\pdfoutput=1
\documentclass[aps,prd,amsmath,floats,floatfix, twocolumn,
superscriptaddress,nofootinbib,showpacs,longbibliography,a4paper]{revtex4-1}

\usepackage[T1]{fontenc}
\usepackage[utf8]{inputenc}
\DeclareUnicodeCharacter{2009}{\nobreakspace}
\usepackage{lmodern}
\usepackage{times}
\usepackage[varg]{txfonts}
\usepackage[normalem]{ulem}
\usepackage{verbatim}
\usepackage[dvipsnames, usenames]{xcolor}
\definecolor{linkcolor}{rgb}{0.0,0.3,0.5}
\usepackage[hypertexnames=false, unicode, colorlinks=true, linkcolor=linkcolor,
citecolor=linkcolor, filecolor=linkcolor,urlcolor=linkcolor,
pdfusetitle]{hyperref}

\usepackage[all]{hypcap}
\usepackage{graphicx}
\usepackage{xspace}
\usepackage{amssymb}
\usepackage{bm} 
\usepackage{microtype}
\usepackage[english]{babel}

\graphicspath{%
  {images/}%
}

\DeclareMathAlphabet{\mathpzc}{OT1}{pzc}{m}{it}

\newcommand{\qeff}{\widetilde{Q}}
\newcommand{\td}{t_{\rm d}}

\begin{document}

\title{Probing charge of compact objects with gravitational microlensing of gravitational waves}
\newcommand{\ICTS}{\affiliation{International Centre for Theoretical Sciences,
    Tata Institute of Fundamental Research, Bangalore 560089, India}}
\newcommand{\IACS}{\affiliation{Indian Association for the Cultivation of
    Science, 2A \& 2B Raja S C Mullick Road, Kolkata 700032, India}}
\newcommand{\IUCAA}{\affiliation{The Inter-University Centre for Astronomy and
    Astrophysics, Post Bag 4, Ganeshkhind, Pune 411007, India}}
\newcommand{\SNU}{\affiliation{Department of Physics and Astronomy,
    Seoul National University, Seoul 08826, Korea}}

\author{Uddeepta Deka}
\email{uddeepta.deka@icts.res.in}
\ICTS
\author{Sumanta Chakraborty}
\email{tpsc@iacs.res.in}
\IACS
\author{Shasvath J. Kapadia}
\email{shasvath.kapadia@iucaa.in}
\IUCAA
\ICTS
\author{Md Arif Shaikh}
\email{arifshaikh.astro@gmail.com}
\SNU
\ICTS
\author{Parameswaran Ajith}
\email{ajith@icts.res.in}
\ICTS

\hypersetup{pdfauthor={Deka et al.}}

\date{\today}

\begin{abstract}
    Gravitational microlensing of gravitational waves (GWs) opens up the exciting possibility of studying the spacetime geometry around the lens. In this work, we investigate the prospects of constraining the `charged' hair of a charged gravitating object from the observation of a GW signal microlensed by the same. The charge can have electromagnetic or modified gravity origin. We compute the analytic form of the lensing potential with charge and construct the lensed waveforms for a range of mass and charge of the charged object, assuming them to be non spinning. Using an approximate likelihood function, we explore how future observations of microlensed GWs can constrain the charge of the lens. We conclude that lensing observations are unlikely to be able to constrain the electromagnetic charge of black holes. However, we might be able to put modest constraints on certain modified gravity models (e.g., the brandworld scenario) or the possibility of the lenses being exotic objects (e.g., naked singularities) . 
\end{abstract}

\maketitle

\section{Introduction}\label{sec:introduction}
The LIGO and Virgo gravitational-wave (GW) detectors have observed $\sim 90$ compact binary coalescence (CBC) events during its first three observing runs (O1, O2, O3) \cite{LIGOScientific:2021djp}. 
Most of them are binary black hole (BBH) mergers. The remaining are mergers of binary neutron stars (BNSs) \cite{LIGOScientific:2017vwq, LIGOScientific:2020aai} and neutron star-black hole (NSBH) binaries \cite{LIGOScientific:2021qlt}. 

These detections have enabled some of the most unique and stringent tests of general relativity (GR) in the strong-field regime \cite{LIGOScientific:2021sio}. These include a model-agnostic residual test that studies the statistical properties of the data after subtracting out the expected GW signal buried therein to ascertain if the residual is consistent with noise \cite{gw150914-tgr, LIGOScientific:2019fpa, LIGOScientific:2020ufj}; an inspiral-merger-ringdown consistency test that checks if the GW waveform is consistent with GR's prediction of the same by comparing the binary's intrinsic parameters inferred from the low- and high-frequency portions of the signal \cite{Ghosh:2016qgn, Ghosh:2017gfp}; a test that compares the speed of GWs with respect to the speed of light \cite{LIGOScientific:2018dkp}, as well as one that probes signatures of velocity dispersion as a consequence of a non-zero graviton mass \cite{will1999}; and a test of GW polarizations that searches for polarizations of GWs beyond the two predicted by GR \cite{LIGOScientific:2017ycc, LIGOScientific:2020tif, Takeda:2020tjj, Wong:2021cmp}.

As with electromagnetic (EM) waves, lensing of GWs results when large agglomerations of matter lead to deviations in the trajectories of these waves \cite{ohanian1974, deguchi1986, Wang1996, Nakamura1998, Bartelmann2010}. The anticipated observations of gravitationally lensed GWs in future observing runs\footnote{No confirmed detection of GW lensing have thus far been reported \cite{hannuksela2019, Dai:2020tpj, McIsaac_2020, LIGOScientific:2021izm, LIGOScientific:2023bwz, Janquart:2023mvf, goyal2023rapid}. Some arguments claiming the observation of lensed GWs -- based on the larger BH masses uncovered by GW observations relative to those inferred from galactic X-ray binaries -- have been made \cite{Broadhurst:2018saj, Broadhurst:2022tjm}.} promise to enable additional novel tests of GR (see, e.g., \cite{Goyal:2020bkm, fan2017, hernandez2022polarization, Finke_2021, narola2023modified, Chung_2021, Goyal_2023, Mukherjee_2020, Collett_2017}), while also shedding light on a number of questions in astrophysics (see, e.g., \cite{Basak:2022fig, Singh:2023hbd, Magare:2023hgs}) and cosmology (see, e.g., \cite{Basak:2021ten, Jana:2022shb, hannuksela2020,Jana:2024dhc}).  

Unlike the case involving EM waves, GW lensing is typically studied in two different regimes. Lensing by galaxies or clusters can be analyzed using the geometric optics (ray-optics) approximation since the wavelengths of the GWs (detectable by LIGO-Virgo) are much smaller than the gravitational radii of such lenses \cite{Ng:2017yiu, Dai:2016igl, Smith:2017mqu}. Strong lensing in this regime results in the production of multiple copies of the GWs \cite{kormann1994, koopmans2009} separated by time delays spanning minutes to months \cite{Haris:2018vmn, More:2021kpb, Magare:2023hgs}. On the other hand, lensing of GWs by massive compact objects with gravitational radii comparable to the GW wavelength will incur wave optics effects \cite{takahashi2003, Jung:2017flg, Diego:2019lcd, Basak:2021ten, urrutia2022}. The result is the production of a single image with a modulated GW shape \cite{Nakamura1998}. This shape carries with it imprints of the properties of the lens, and can possibly be used to probe the nature of the compact lens.

Wave-optics effects modulating the GW waveform have been studied extensively for simple lensing potential models, in particular the one corresponding to the point mass lens\footnote{Most searches for microlensing signatures incurred by wave-optics effects in detected GW events assume a point-mass lens model \cite{Basak:2021ten, hannuksela2019, LIGOScientific:2023bwz}.} \cite{Nakamura1998, Nakamura1999, takahashi2003}. In this model, the frequency-dependent amplitude modulations of the GW waveform are exclusively determined by the (redshifted) mass of the lens, and the source position (i.e, the location of the source in the lens plane) \cite{takahashi2003}. However, if the compact object has additional hairs, they are also going to show up in the lensed GW waveform, and can possibly be detected as and when microlensed GW signals are observed. The additional hairs could point to the specific nature of the compact object or point to a deviation from GR. 

Besides the mass, the next obvious hair that a static and spherically symmetric compact object can have is the electric charge $q$, which modifies the radial and temporal components of the spacetime metric by $\mathcal{O}(\mathcal{Q}/r^2)$, where $\mathcal{Q}\equiv q^{2}$. Indeed, it is widely believed that astrophysical objects should not have any net electric charge, or, even if there is some electric charge, it should get shielded \cite{Feng:2022evy, Pina:2022dye, Bozzola:2020mjx} or neutralised~\cite{Eardley:1975kp}. Intriguingly, such a `charge-like' hair can arise from several other contexts as well and can have non-negligible values. For example, in the context of an extra spatial dimension, where we live in a four-dimensional universe embedded in a five-dimensional spacetime, known as the braneworld scenario, the solution of the effective gravitational field equations on the four-dimensional brane has exactly the same structure, but with $\mathcal{Q}$ being \emph{negative} \cite{Shiromizu:1999wj, Dadhich:2000am, Harko:2004ui, Aliev:2005bi, Maeda:2006hj}. Similarly, a positive value of $\mathcal{Q}$ can arise in scalar coupled Maxwell theories, described by Horndeski theories \cite{Babichev:2015rva, Barrientos:2017utp, Babichev:2013cya}. There have already been extensive searches for both positive as well as negative values of the `charge' in various contexts ---  (a) in the weak field regime, e.g., through solar system tests \cite{Iorio:2012cm, Capozziello:2012zj, Chakraborty:2012sd, Bhattacharya:2016naa, Mukherjee:2017fqz}, (b) using EM waves from accretion on supermassive BHs \cite{Maselli:2014fca, Stuchlik:2008fy, Banerjee:2021aln, Banerjee:2017hzw}, (c) using GWs from BBH and BNS coalescence \cite{Barausse:2015wia, Toshmatov:2016bsb, Andriot:2017oaz, Chakraborty:2017qve, Chakravarti:2019aup, Mishra:2021waw, Mishra:2023kng, Gupta_2021, Carullo:2021oxn, gu2023constraints}, and (d) with strong-field lensing of EM waves and measurement of BH shadows \cite{Chakraborty:2016lxo, Banerjee:2022jog, Vagnozzi:2022moj, Banerjee:2019nnj}.     

Here we explore the possibility of constraining the charge hair of an astrophysical object from the observation of GWs microlensed by it. We study the lensing-induced modulations on the GW signal by introducing the charge in the lensing potential. In particular, we model the spacetime surrounding the lens to be described by a metric that is analogous to the Reisner-Nordstr\"{o}m spacetime, with $\pm(\mathcal{Q}/r^{2})$ term in the temporal and radial metric elements. If a lensed GW signal suggests a positive value of $\mathcal{Q}$, it is to be interpreted either in the context of EM-charged object, or, as an astrophysical object in modified theories of gravity. While, if the lensed GW signal prefers a negative value for $\mathcal{Q}$, then it will provide a hint for the existence of extra dimensions. In what follows we will compute the lensing potential associated with the new metric, and numerically calculate the corresponding frequency-dependent magnification function that modulates the GW, for a range of the lens masses, source positions (i.e, the angular location of the source in the image plane), and the charge parameter $\mathcal{Q}$. 

To assess our ability to constrain the charge $\mathcal{Q}$, we quantify the extent to which the GW signal lensed by an object of charge $\mathcal{Q}$ deviates from another waveform lensed by another object with charge $\mathcal{Q}_\mathrm{true}$. To that end, we compute matches, defined as the inner product between two waveforms weighted by the noise power spectral density (PSD) of the GW detector. We then use these to construct an approximate likelihood on the lens parameters, which we sample to assess how well we can recover the true value of the charge. 


We find that the correction to the point-mass lens potential due to a charge $\mathcal{Q}$ is given by the effective parameter $\qeff\equiv {3\pi \mathcal{Q}}/{16\xi_{0}M_{\rm L}}$, where $M_L$ is the mass and $\xi_0$ is the Einstein radius of the lens. We find that when the lens is actually chargeless, we will be able to constrain the the charge parameter to be $|\qeff| \lesssim 10^{-3}$. Note that, in typical astrophysical situations, $\xi_0 \sim (M_L D_L)^{1/2}$, where $D_L$ is the distance to the lens. Thus, the expected constraints on the dimensionless charge $\mathcal{Q}/M_L^2$ are $\mathcal{O}(10 ^{-3} \sqrt{D_L/M_L})$. Indeed, our prospective constraints are much worse than the maximum possible charge of a BH ($\mathcal{Q}/M_L^2 = 1$). Thus, the prospects of measuring the electric charge of a BH using GW lensing are grim. However, these bounds can constrain the possibility of the lens being an exotic compact object (like a naked singularity) and can constrain some of the alternative theories of gravity, which predict large values of $\mathcal{Q}$.

The paper is organized as follows: In Section \ref{sec:amplification_due_to_a_charged_lens}, we derive the analytical expression for the lensing potential in the presence of the charge $\mathcal{Q}$. Using the expression for the lensing potential, in Section \ref{sec:computing_the_amplification_factor} we numerically compute the magnification function, as well as the GW waveform lensed by a charged object. In Section \ref{sec:results} we explore the possibility of constraining the charge parameter from the observation of GWs microlensed by it. We conclude in Section \ref{sec:conclusion}. Some of the detailed calculations are presented in Appendix \ref{appna}. 

\emph{Notations and Conventions:} We will set the fundamental constants {$G$, $c$ and $1/(4\pi\epsilon_0)$} to unity and will use the mostly positive signature convention, such that the Minkowski metric in Cartesian coordinate takes the form $\eta_{\mu \nu}=\textrm{diag}(-1,+1,+1,+1)$. Greek indices $\mu, \nu, \cdots$ denote four-dimensional spacetime coordinates, while uppercase Roman indices $A, B, \cdots$ denote five-dimensional spacetime indices. We will refer to $\mathcal{Q}$ as the `charge', or `charge hair' of a BH. Note that, even a negative electric charge $q$ will correspond to a positive value of $\mathcal{Q}$, since $\mathcal{Q}\equiv q^{2}$. Also, we will refer to lensing in the wave optics regime ($G M_\mathrm{L}/c^2 \sim \lambda_\mathrm{GW}$) as microlensing.

\section{Lensing magnification by a charged black hole}
\label{sec:amplification_due_to_a_charged_lens}
In this section, we will determine the magnification function due to the simplest possible hair of an astrophysical object, namely a charge that can either have an EM origin or, may arise from theories of gravity beyond GR. Of course, any EM charge is expected to be dissipated away with time, or, would be heavily shielded by surrounding matter. But it would still be interesting \cite{Feng:2022evy, Pina:2022dye, Bozzola:2020mjx} if any residual electric charge present in an astrophysical object can be tested using gravitational lensing. More interesting is the case when this charge arises from an alternative theory of gravity. We will provide some examples of such modified theories below. 

First of all, the presence of an extra spatial dimension can induce such a charge in the four-dimensional spacetime metric outside of a BH. This is because the effective gravitational field equations on the four-dimensional vacuum spacetime (known as the brane) take the following form \cite{Shiromizu:1999wj, Dadhich:2000am}
\begin{align}
G_{\mu \nu}+E_{\mu \nu}=0~.
\end{align}
Here $G_{\mu \nu}$ is the four-dimensional Einstein tensor and $E_{\mu \nu}=W_{ABCD}e^{A}_{\mu} n^{B} e^{C}_{\nu}n^{D}$ is the projected five dimensional Weyl tensor on the brane, with $e^{A}_{\mu}$ being the projector and $n_{A}$ being the normal to the brane. Owing to the symmetries of the Weyl tensor, it follows that $E^{\mu}_{\mu}=0$, which is akin to the traceless property of the energy-momentum tensor of the EM field. Thus the above field equation for the metric admits the following static and spherically symmetric solution, $-g_{tt}=g^{rr}=1-(2M/r)+(\mathcal{Q}/r^{2})$, with $\mathcal{Q}<0$ and thus differing from the Reissner-Nordstr\"{o}m solution by the sign of the $(1/r^{2})$ term. Note that, in the system of units we are considering, the charge $\mathcal{Q}$, as experienced by a three-dimensional observer, has a dimension of $\textrm{(Length)}^{2}$. In the $f(T)$ gravity as well the same solution for the metric components was obtained \cite{Capozziello:2012zj}, however, with a positive contribution from the charge term $\mathcal{Q}$. The Einstein-Gauss-Bonnet theory in higher dimension also admits the same solution on the brane with $\mathcal{Q}<0$, albeit the origin of the charge term was from the coupling constant in the Gauss-Bonnet invariant \cite{Maeda:2006hj}. Finally, the same solution also arises in the context of a sub-class of Horndeski theories of gravity, which involves the following terms: $\beta G^{\mu \nu}\partial_{\mu}\phi \partial_{\nu}\phi$, as well as $-(\gamma/2)T^{\mu \nu}\partial_{\mu}\phi \partial_{\nu}\phi$ in the gravitational action, besides the Ricci scalar and the Maxwell term. In this case, the charge $\mathcal{Q}$ depends on the ratio $(\gamma/\beta)$ and is a positive quantity \cite{Babichev:2015rva}. Thus, we observe that various cases yield the same metric, with different origin and sign for the charge term. In summary, there are several possibilities for a positive charge to appear in the metric --- (a) EM charge, (b) charge arising from the $f(T)$ theory of gravity, (c) charge depending on the non-trivial coupling between gravity and electromagnetism with scalar, as in Hornsdeski theories. While the negative charge predominantly arises in the presence of higher dimensions, either from the bulk Weyl tensor or, from the Gauss-Bonnet coupling. This makes the charged metric an ideal ground to probe, since it arises in so many different contexts, with many varieties and hence it is worthwhile to see the effect of such a charge on the microlensing of GWs. In particular, when the wavelength of GWs is comparable to the radius of the event horizon of these charged BHs wave effects must be taken into account. In what follows, we compute the effect of this charge term on the amplification of GWs due to microlensing. 

\begin{figure*}[t]
      \centering
      \includegraphics[width=\textwidth]{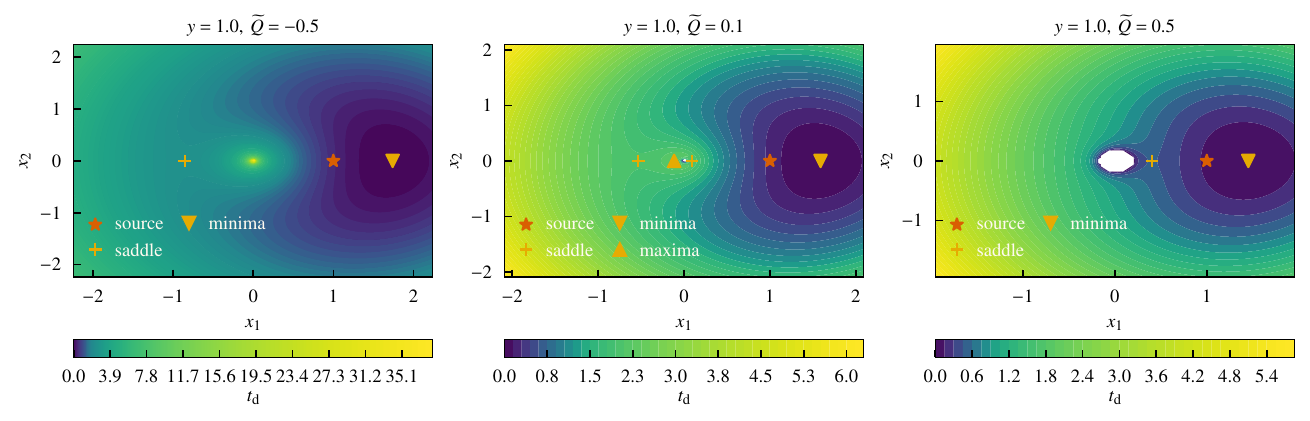}
      \caption{The time delay contours, $\td$ {(in units of $4 M_{\rm L} (1+z_{\rm L})$ and $t_{\rm d}=0$ at the global minimum)} plotted as a function of the lens plane coordinates $(x_{1},x_{2})$. The source position is shown by a star ($y = 1.0$). The left, center, and right panels correspond to the cases with $\qeff=-0.5,\, 0.1\text{ and } 0.5$, respectively. For each of these cases, the locations of the images and their types are also indicated.}
      \label{fig:td_map}
\end{figure*}

We will work within the thin lens approximation, i.e., the lens is not considered a three-dimensional object, but rather a two-dimensional one. This is because the size of the lens is much smaller compared to the distance of the lens to the source as well as to the observer. Thus the magnification function of the GW, as measured by an observer reads \cite{takahashi2003},
\begin{align}
\mathcal{F}(f)=\frac{D_{\rm S}\xi_{0}^{2}}{D_{\rm L}D_{\rm LS}}\frac{f}{i}\int d^{2}\vec{x}\exp\left[2\pi if t_{\rm d}(\vec{x},\vec{y})\right]~,
\end{align}
where $\vec{x}\equiv \vec{\xi}/\xi_{0}$ is the normalized vector on the lens plane and $\vec{y}=(D_{\rm L}/\xi_{0}D_{\rm S})\vec{\eta}$ is the normalized vector on the source plane indicating the location of the source. Further, $D_{\rm L}$ is the angular diameter distance to the lens, $D_{\rm S}$ is the angular diameter distance to the source and $D_{\rm LS}$ is the angular diameter distance between the source and the lens. Further, the quantity $t_{\rm d}$ is the time taken by the GW to reach the observer from the source when a lens is present, while $f$ is the frequency of the GW emitted from the source. Since the source and the observer, in general, are separated by cosmological distances, the effect of the cosmological expansion must be included in the above analysis. This modifies the above magnification function to,
\begin{align}\label{eq:ampfac_integral_form}
\mathcal{F}(f)&=\frac{D_{\rm S}\xi_{0}^{2}\left(1+z_{\rm L}\right)}{D_{\rm L}D_{\rm LS}}\frac{f}{i}
\int d^{2}\vec{x}\exp\left[2\pi ift_{\rm d}(\vec{x},\vec{y})\right]~,
\end{align}
where, $z_{\rm L}$ is the redshift of the lens. Note that the magnification function $\mathcal{F}(f)$ is so chosen that $|\mathcal{F}(f)|=1$ in the absence of the lens. 
{The lensing time delay $t_{\rm d}(\vec{x},\vec{y})$ (as compared to an unlensed signal) is due to two contributions: 1) the longer geometric path taken by the wavefront due to the deflection, and 2) the gravitational time delay due to the lensing potential.
\begin{align}\label{eq:time_delay_raw}
t_{\rm d}(\vec{x},\vec{y}) =  \frac{D_{\rm S}\xi_{0}^{2}\left(1+z_{\rm L}\right)}{D_{\rm L}D_{\rm LS}} ~ \left[ \frac{|\vec{x}-\vec{y}|^{2}}{2}-\psi(\vec{x}) \right].
\end{align}
Here, $\psi(\vec{x})$ is the two-dimensional deflection potential, which is obtained by projecting the three dimensional potential $\Phi$  on to the lens plane. 
(i.e., perpendicular to the line of sight coordinate $z$, connecting observer and the lens). Here, $\Phi$ is the effective gravitational potential, not to be confused with the Newtonian potential, which is the $g_{tt}$ component of the metric (see Appendix~\ref{appna} for the derivation)
\begin{equation}
\label{eq:effective_newtonian_potential_main}
\Phi = \frac{-GM_{\rm L}}{R} + \frac{3G\mathcal{Q}}{8R^2}\,.
\end{equation}
Thus, the projected potential is 
\begin{align}
\psi(\vec{x})=\ln \mid \vec{x} \mid + \frac{\qeff}{\mid \vec{x}\mid }~, \mathrm{where}~~~  \qeff\equiv \frac{3\pi \mathcal{Q}}{16\xi_{0}M_{\rm L}}~.
\label{eq:lens_potential_charged}
\end{align}

Here, we have assumed $\xi_{0}^{2}=(4M_{\rm L}D_{\rm L}D_{\rm LS}/D_{\rm S})$, to be the Einstein radius. Given that the charge $\mathcal{Q}$ has the dimension of $(\textrm{length})^{2}$, it follows that $\qeff$ is dimensionless, which is consistent with the fact that the lensing potential and $\vec{x}$ are dimensionless. Note that in the limit $\mathcal{Q}\to 0$, we reproduce the expression of the point mass lens.}  

Given this gravitational potential, in the geometric optics regime, the location of the images on the lens plane can be determined by invoking Fermat's principle of gravitational lensing~\cite{1986ApJ...310..568B}. Fermat's principle states that the images are formed at the extrema of the time delay surface, that is, at points on the lens plane satisfying the condition: 
\begin{align}\label{image_location_condition}
\frac{\partial t_{\rm d}(\vec{x},\vec{y})}{\partial \vec{x}} = 0. 
\end{align}
This results in the \emph{lens equation}
\begin{equation}\label{image_location}
\vec{y} = \vec{x} - \left(1-\frac{\qeff}{\mid\vec{x}\mid}\right)\frac{\vec{x}}{\mid \vec{x}\mid^2}, 
\end{equation}
which can be solved to obtain the image locations. Note that GW microlensing happens in the wave optics regime, where the geometric approximation is not valid. However, an approximate understanding of the image locations is useful in interpreting the lensing magnification that we compute in the wave optics regime. 

In the case of the point mass (uncharged) lens, the diffraction integral in Eq.\eqref{eq:ampfac_integral_form} can be solved analytically, as in \cite{takahashi2003}. However, in the presence of charge, the scalar wave equation becomes too complicated to render an analytical solution possible (see Appendix \ref{appna}). Nevertheless, Eq.\eqref{eq:ampfac_integral_form} can be solved numerically using the lensing potential Eq.\eqref{eq:lens_potential_charged} as the essential features of the diffraction integral remain unaltered.

\begin{figure}[t]
      \centering
      \includegraphics{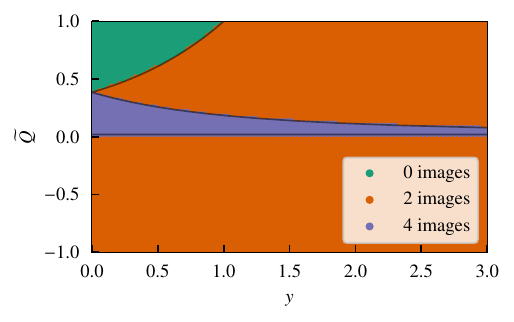}
      \caption{Number of images shown as a function of the source position $|\vec{y}|$, and the effective charge $\qeff$. There exist two images for all $\qeff\leq 0$, irrespective of the source position. However, for $\qeff>0$, there are three possibilities - zero, two or four images. See text for discussion.}
      \label{fig:num_images}
\end{figure}
It turns out that for $\mathcal{Q}\leq 0$, the lens equation Eq.\eqref{image_location} has two solutions for $\vec{x}$. These will be the image locations in the geometric optics limit. One of these solutions corresponds to the saddle point of the time delay surface, while the other is the minima, as depicted in the left panel of Fig. \ref{fig:td_map}. Besides the locations of these two points, Fig. \ref{fig:td_map} shows the contours of constant $t_{\rm d}(\vec{x},\vec{y})$ for a fixed source location $\vec{y}$. Interestingly, for certain choices of $\mathcal{Q}>0$, and the source position $|\vec{y}|$, there can even be four solutions to Eq. \eqref{image_location} for $\vec{x}$, leading to the formation of four images. This can be seen in Fig. \ref{fig:num_images} and the middle panel in Fig. \ref{fig:td_map}. Finally, for large value of $\mathcal{Q}$, it follows that there are no images, this is because, the images are just behind the lens and hence unresolvable.

The formation of multiple images can be understood using caustics associated with the lensing potential. For $\mathcal{Q}\leq 0$, there are no caustics {(except for the lens location)} and hence there are always two images. For any $\mathcal{Q}>0$, a caustic appears at a finite value of $|\vec{y}|$, which decreases as the charge increases. Therefore, for $\mathcal{Q}$ smaller than certain value, the source will appear within the caustic, leading to four images. As the charge increases, the caustic appears at smaller values of $|\vec{y}|$, and for $\mathcal{Q}$ larger than certain value, the caustic reaches $|\vec{y}|\sim 0$. Thus the source goes out of the caustic, leading to two images.  
For even higher value of $\mathcal{Q}$, it follows that the image appears just behind the source and hence the image is unresolvable. This is precisely what Fig. \ref{fig:num_images} demonstrates.

Note that the above situation cannot be compared with the scenario involving the BH shadow, which involves the lensing of EM waves in the strong gravity regime. There, the source (emission from the accreting gas) is close to the lens and the light rays {needs to be traced using the full BH metric.}
In contrast, typical gravitational lensing (as considered here) considers the weak field regime. Also, the distance between the source and the lens is so large that the lens can be approximated by a projected two-dimensional potential (thin lens approximation). Hence the details of the images are quantitatively different in the two scenarios. {However, as expected, the lensing potential derived above, gives rise to a deflection angle, which in the geometric optics regime, matches with the weak field limit of the strong field calculations presented in \cite{Eiroa:2002mk,Eiroa:2014mca,Eiroa:2003jf} (see Appendix \ref{appna})}.

{Similar to the case of the point mass lens, the images and the source appear along a line in the plane spanned by $\vec{y}$ (see Fig.~\ref{fig:td_map}), which arises from the axial symmetry of the lens}. This is also reflected in the fact that the lensing potential $\psi(\vec{x})$ depends on $|\vec{x}|$ alone. Thus it follows that the solution of Eq. \eqref{image_location} depends on $|\vec{y}|$ alone, and hence throughout the rest of the analysis we will denote $y\equiv |\vec{y}|$, which is the only input required for determining the image location and the associated time delay contours. 


Some comments regarding the values of $\widetilde{Q}$ are in order. Note that, except for an overall normalization factor the effective charge $\widetilde{Q}$ can be written as, $\qeff \simeq (\mathcal{Q}/M_{\rm L}^{2})(M_{\rm L}D_{\rm S}/D_{\rm L}D_{\rm LS})^{1/2} \sim (\mathcal{Q}/M_{\rm L}^{2})(M_{\rm L}/D_{\rm L})^{1/2}$. Thus, even in the extremal limit of BHs ($\mathcal{Q}/M_{\rm L}^{2}=1$), the effective charge is $\qeff \simeq (M_{\rm L}D_{\rm S}/D_{\rm L}D_{\rm LS})^{1/2} \sim (M_{\rm L}/D_{\rm L})^{1/2} \ll 1$. This is because, the typical size of the BH is much smaller than any other length scale in the problem, e.g., the source-lens, or, the lens-observer distance. Thus for a typical positive value of $\qeff$ that we can constrain ($\qeff \sim 10^{-3}$), the compact object describes a naked singularity. On the other hand, for negative values of $\qeff$, there is always a horizon and hence it depicts a BH.

\section{Numerical computation of the lensing magnification}
\label{sec:computing_the_amplification_factor}
\begin{figure*}[t]
    \centering
    \includegraphics[width=\textwidth]{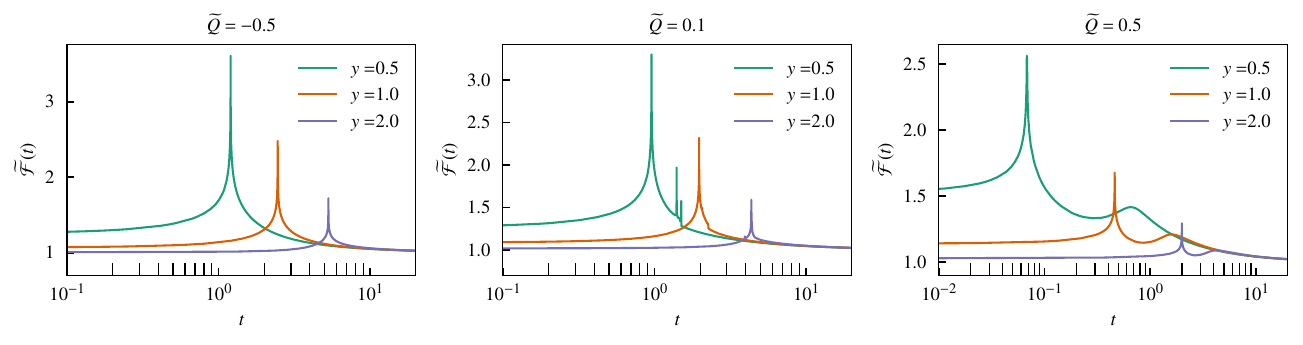}
    \caption{The time domain magnification function $\widetilde{\mathcal{F}}(t)$ as a function of time $t$ ({in units of $4M_{\rm L} (1+z_{\rm L})$}, where the global minima lies at $t=0$). The panels from left to right correspond to cases with $\qeff=-0.5,\, 0.1\text{ and }0.5$, respectively. The plots in each of these panels correspond to source position values $y=0.5, \,1.0 \text{ and }2.0$. In the left and right panels, there are always two images -- a minima image and a saddle image. The minima image lies at $t=0$ while the saddle image can be identified by the logarithmically diverging peak. On the center panel, there are four images -- one minima image, two saddle images, and one maxima image.}
    \label{fig:F(t)}
\end{figure*}

In this section, we will determine the magnification function due to a charged lens and will find out how the GW waveform is modulated due to lensing. For an isolated point mass lens, the lensing magnification function $\mathcal{F}(f)$ can be derived analytically~\cite{Nakamura1999}. Unfortunately, for the present case of a charged lens, an analytical expression for the magnification function cannot be obtained and one must resort to numerical schemes. Implementing such numerical schemes is also difficult in the present context owing to the highly oscillatory behaviour of the integral, which renders the traditional numerical integration methods to be computationally ineffective. Therefore, we developed and implemented a new numerical scheme to compute the magnification function, which we outline below. This is an extension of the methods used in~\cite{Diego:2019lcd}, based on the original work by \cite{Ulmer_1995}. The new ingredient is to use a histogram for an efficient computation of the diffraction integral Eq.\eqref{eq:diff_int_td} in time domain.
\begin{enumerate}

\item \textsc{Fixing the length scale} $\xi_{0}$: The time delay contours, as well as the lensing potential, depend on the length scale $\xi_{0}$, which we fix to be of the following form:
\begin{equation}\label{eqn:einstein_radius}
\xi_0^2=4M_0\frac{D_{\rm L}D_{\rm{LS}}}{D_{\rm S}}~,
\end{equation}
where $M_0$ is an arbitrary mass scale. In the present problem, the only relevant mass scale is set by the lens mass and hence we choose $M_0=M_{\rm L}$. With this choice of the length scale $\xi_{0}$, Eq. \eqref{eq:time_delay_raw} becomes,
\begin{align}
t_{\rm d}(\vec{x},\vec{y})&=4M_{\rm L}(1+z_{\rm L})
~ \left[\frac{|\vec{x}-\vec{y}|^{2}}{2}-\psi(\vec{x})\right]~.
\end{align}
In the subsequent analysis, we will use the above expression for the time delay.  


\begin{figure*}[t]
    \centering
    \includegraphics{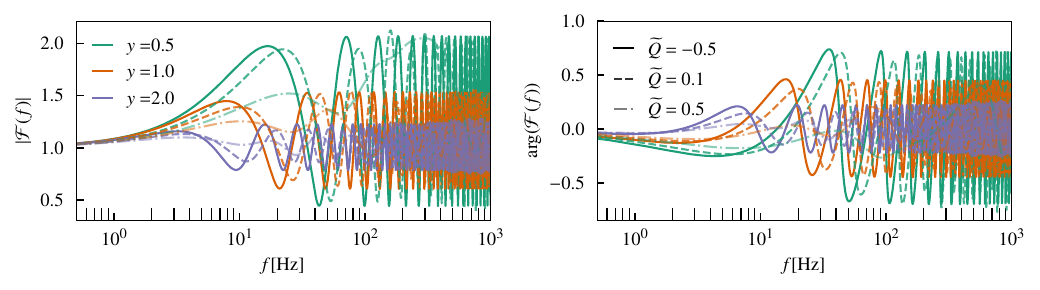}
    \caption{The frequency domain magnification function $\mathcal{F}(f)$ as a function of the GW frequency (in Hz) for a lens with mass $M_{\rm L}=500M_\odot$, at redshift $z_{\rm L}=0.5$. The left panel corresponds to the amplitude of $\mathcal{F}(f)$, while the right panel corresponds to its phase. The plots correspond to cases with $\qeff=-0.5 $ (solid), $\qeff=0.1$ (dashed) and $\qeff=0.5$ (dashed-dotted), for source position values $y = 0.5, 1.0 \text{ and }2.0$ (plots with different colors).}
    \label{fig:F(w)}
\end{figure*}

\item \textsc{Time domain magnification function}: As we have already discussed, determining the magnification function $\mathcal{F}(f)$ in the frequency domain requires performing the highly oscillatory integral over the time function $t_{\rm d}(\vec{x},\vec{y})$, which is numerically challenging. Thus, we will first compute the magnification function in time domain, and then compute $\mathcal{F}(f)$ using Fourier transform. For this purpose we rewrite Eq. \eqref{eq:ampfac_integral_form} in the following form:
\begin{equation}
\mathcal{F}(f) = C(f)\int d^2\vec{x}\,\exp[2\pi i f t_{\rm d}(\vec{x}, \vec{y})]~.
\end{equation}
We define the time domain magnification function $\widetilde{\mathcal{F}}(t)$ as the inverse Fourier transform of the ratio $\{\mathcal{F}(f)/C(f)\}$, which yields:
\begin{align}
\widetilde{\mathcal{F}}(t)&\equiv \int df\,\frac{\mathcal{F}(f)}{C(f)}\,\exp[-2 \pi if t]
\nonumber
\\
&= \int d^2\vec{x}\,\delta[t_{\rm d}(\vec{x}, \vec{y})-t]~.
\label{eq:diff_int_td}
\end{align}
As evident from the above expression, the time-domain magnification function is related to the area between the time-domain contours in the lens plane. In particular, $\widetilde{\mathcal{F}}(t)dt$ is the area $dS$ between the curves $t$ and $t+dt$ of constant time delay, i.e., 
\begin{equation}
\widetilde{\mathcal{F}}(t)=\frac{dS}{dt}~.
\end{equation}
Thus by computing the area between two infinitesimal time domain contours, we can determine the magnification function $\widetilde{\mathcal{F}}(t)$ in the time domain. 

\item \textsc{Histogram of the time delay map}: To find the area between the contours of constant time delay, we place a uniform grid on the lens plane and determine the time delay at each of the grid points (see, e.g., Fig.~\ref{fig:td_map}). This yields the number of grid points between time delays $t$ and $t+dt$, which leads to a histogram for the number of grid points at each of the time delay contours $t$, with bin size $dt$. This number is a proxy for the area between the time delays $t$ and $t+dt$. Thereby allowing us to compute the time-domain magnification function $\widetilde{\mathcal{F}}(t)$.

For a charged lens, the above method of determining the area between time delay contours leads to the desired time-domain magnification function  (Fig.~\ref{fig:F(t)}). As evident, the peak of the amplification is higher for $\qeff<0$ and for smaller but positive $\qeff$, while it decreases for large, positive $\qeff$. The peak is also higher for smaller values of the source position $y$, since the GW passes closer to the lens and is most affected. Another intriguing feature, already noted earlier, can also be seen in Fig. \ref{fig:F(t)}, namely, the existence of four images for certain positive values of $\qeff$ and for specific values of $y$. For $\qeff=0.1$ and $y=0.5$, these four images can be seen explicitly. The one at $t=0$ is present in all the plots, which have not been shown, then two additional images are clearly visible, while the fourth one (maxima image) is very close to one of the saddle images. For larger values of $y$, there are two images with positive $\qeff$, but some additional bumps are present in the time domain magnification (right panel of Fig.~\ref{fig:F(t)}).

\item \textsc{Frequency domain magnification function}: Having computed the time-domain magnification function $\widetilde{\mathcal{F}}(t)$, it is straightforward to determine the frequency domain magnification $\mathcal{F}(f)$. For this purpose, we simply perform a Fast Fourier Transform (FFT) of $\widetilde{\mathcal{F}}(t)$ and multiply by the overall factor $C(f)$, yielding:
\begin{equation}
\mathcal{F}(f)=C(f)\times \mathrm{FFT}\left[\widetilde{\mathcal{F}}(t)\right]~.
\end{equation}
Though the time-domain magnification function was real, the frequency-domain amplification is complex. Following this, we have plotted the magnitude $|\mathcal{F}(f)|$ and the argument $\textrm{arg}[\mathcal{F}(f)]$ in Fig. \ref{fig:F(w)}. As expected, the frequency-domain magnification also is the largest for $\qeff<0$, and smaller positive $\qeff$, while is the smallest for larger and positive values of $\qeff$. The magnification function decreases as the source position increases.  

  \begin{figure*}[t]
    \centering
    \includegraphics[width=\textwidth]{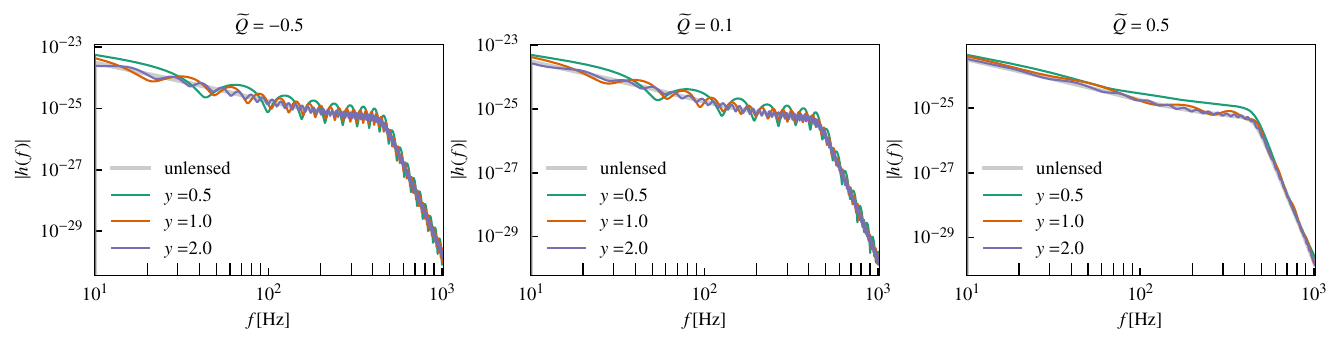}
    \includegraphics[width=\textwidth]{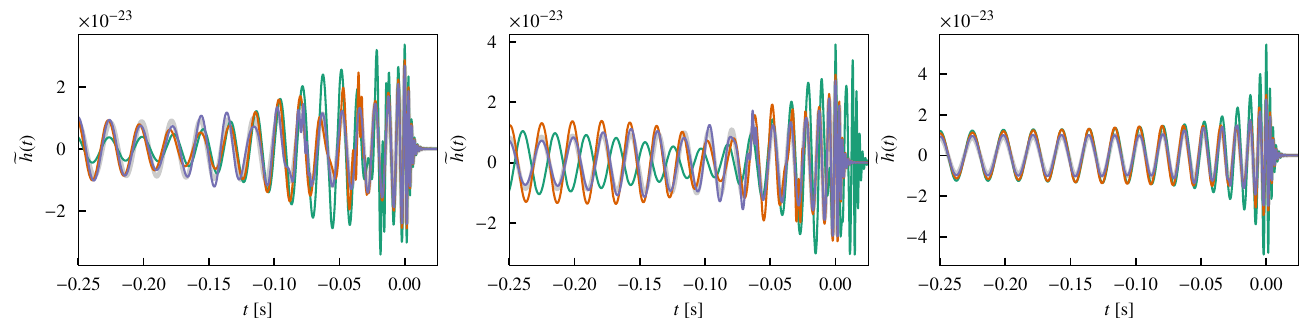}
    \caption{GW signals lensed by charged BHs of mass $M_{\rm L}=500M_\odot$, at redshift $z_{\rm L}=0.5$. The panels from the left to the right correspond to cases with $\qeff=-0.5,\, 0.1\text{ and }0.5$ respectively, for various source positions (shown in legends). The top and bottom panels show the amplitude of lensed waveforms in the frequency and time domains, respectively. The solid grey lines indicate the unlensed waveforms.}
    \label{fig:lensed_waveforms}
  \end{figure*}

\item \textsc{Lensed waveform}: Finally, the GW strain is going to be amplified by the frequency-dependent magnification factor. In the frequency domain, the unlensed GW strain $h^U(f|\vec{\theta})$ is going to be modified as:
\begin{equation}\label{eq:hLensed}
h^{L}(f; \vec{\theta}, \vec{\lambda})= \mathcal{F}(f; \vec{\lambda}) ~ h^U(f; \vec{\theta}) ~.
\end{equation}
Here $h^L(f; \vec{\theta}, \vec{\lambda})$ is the lensed waveform, $\vec\theta$ are the source parameters of the GW signal and $\vec\lambda=(M_{\rm L}^z,y,\qeff)$ are the lens parameters, where $M_{\rm L}^z \equiv M_{\rm L}(1+z_{\rm L})$ is the redshifted lens mass. The GW signal lensed by a charged lens has been shown in both the frequency and the time domain in Fig. \ref{fig:lensed_waveforms}, along with the corresponding unlensed waveform. It is evident (in particular, from the time domain plots) that the lensed waveform for $\qeff<0$, as well as for $\qeff>0$, with four images, has the largest departure, while for positive $\qeff$, with two or no images, has the least departure from the unlensed waveform. For $\qeff<0$, the lensed and unlensed waveforms are mostly in phase, while for $\qeff>0$, with four images, these waveforms are out of phase. Again for positive $\qeff$ with two images, the lensed waveform is mostly in phase. As we will see, these properties make constraining negative as well as small positive values of $\qeff$ much easier, than constraining large positive values of $\qeff$.  

\end{enumerate}

As a verification of the numerical scheme described above, we have employed the above method for determining the lensed waveform for a point mass (uncharged) lens. Since the magnification function for a point mass lens can be computed analytically~\cite{Nakamura1998}, we can compare it with our numerical results. Following this, we have evaluated the mismatch between the lensed waveforms computed using the analytical form of the magnification function for isolated point mass lens and lensed waveforms computed using our method with $\qeff=0$. We find that the mismatch values are within permissible limits ($\lesssim 10^{-5}$). This validates the numerical method described above and has been used for determining the magnification function for a charged lens.

\section{Prospective constraints on the black hole charge from lensing observations}\label{sec:results} 
In this section, we will derive prospective constraints on the BH charge from future observations of microlensed GW signals. We have already noticed that both negative and positive values of the charge $\qeff$ significantly modify the GW waveform through lensing and the effect is present in the amplitude, as well as in the phase of the GW signal (Fig.~\ref{fig:lensed_waveforms}). Thus, we expect that a BH lens without charge can be efficiently distinguished from a charged BH lens for $\qeff<0$, as well as for small positive values of $\qeff$, while we expect such a distinction to be less efficient for larger and positive values $\qeff$. 

To get a better understanding of the prospective constraints on BH charge from future observations of microlensed GWs, we estimate the (approximate) Bayesian posteriors of $\qeff$ from simulated observations of microlensed GWs. The GW signal microlensed by a BH can be parameterised by a set of source parameters $\vec{\theta}$ and lens parameters $\vec{\lambda}$. Their posterior distributions can be estimated from the observed data $d$ using the Bayes theorem: 
\begin{equation}
p(\vec{\theta},\vec{\lambda}|d) = \frac{p(\vec{\theta},\vec{\lambda}) p(d|\vec{\theta},\vec{\lambda},\mathcal{H}_\mathrm{ML})}{p(d|\mathcal{H}_\mathrm{ML})}~,
\end{equation}
where $p(\vec{\theta},\vec{\lambda})$ is the prior distribution of the source and lens parameters, $p(d|\vec{\theta},\vec{\lambda},\mathcal{H}_\mathrm{ML})$ is the likelihood of getting data $d$ given the parameters  $\vec{\theta},\vec{\lambda}$ and the hypothesis $\mathcal{H}_\mathrm{ML}$ that the data contains a microlensed GW signal. The denominator,  $p(d|\mathcal{H}_\mathrm{ML})$, is the likelihood marginalised over all the parameters (called, the evidence of the hypothesis $\mathcal{H}_\mathrm{ML}$). 

%
\begin{figure*}[t]
\centering\
\includegraphics[width=6in]{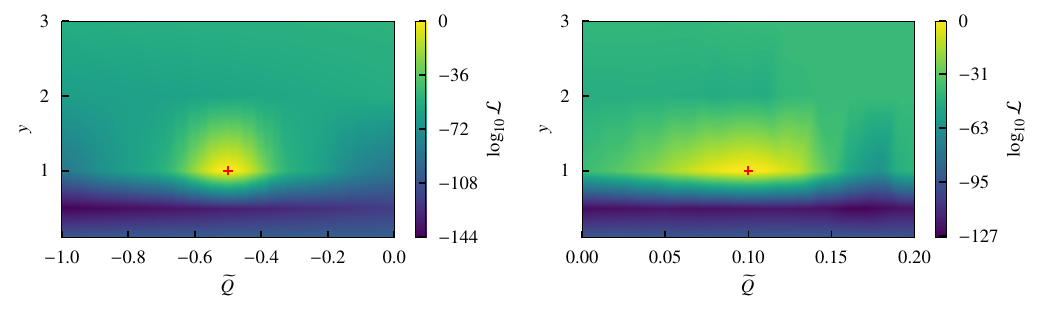}
\includegraphics[width=6in]{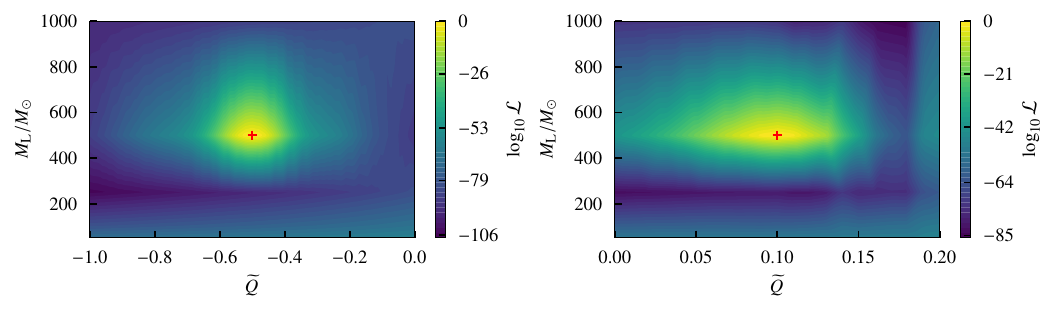}
\caption{The top panel shows the joint likelihood in the source position $y$ and effective charge $\qeff$ space, for a GW signal with $\rm{SNR}=25$ microlensed by a charged BH of mass $M_{\rm L}=500M_\odot$. The true values of the source position $(y=1)$ and charge ($\qeff = -0.5, 0.1$) are indicated by the red markers. The bottom panel shows the joint likelihood in the $M_{\rm L}/M_\odot$ and $\qeff$ space, for a GW signal with $\rm{SNR}=25$ microlensed by a charged BH with  $y = 1$. The true values of the lens mass $(M_{\rm L}=500M_\odot)$ and charge ($\qeff = -0.5, 0.1$) are indicated by the red markers. It can be seen that $\qeff$ is largely uncorrelated with other parameters $M_{\rm L}$ and $y$.}
\label{fig:2D_lkhd_y_Q}
\end{figure*}

The posterior $p(\qeff|d)$ of the charge can be computed by marginalising $p(\vec{\theta},\vec{\lambda}|d)$ over all parameters except $\qeff$. For simplicity, we will assume that the lens parameters $\vec{\lambda}$ are largely uncorrelated with the source parameters $\vec{\theta}$. Hence we need to compute the likelihood only on $\vec{\lambda}$ for estimating prospective constraints on $\qeff$. This is a reasonable assumption 
although recent work has identified possible correlations between microlensing modulations and modulations induced by spin-induced precession~{\cite{Mishra:2023ddt}}. An uncharged point mass lens in the background of a macro lens (e.g., a galaxy) could also introduce more complex modulations in the GW signal~\cite{Meena:2019ate,Diego:2019lcd,Cheung:2020okf,Mishra:2021xzz}, potentially mimicking some of the effects of a charged lens. Also, note that currently, we assume non-spining BH lenses.  There could be some correlations between the charge and the spin of a BH~(see, e.g.,~\cite{Carullo:2021oxn}). For the time being, we ignore these additional complexities. 

It turns out that the lens parameters $M_\mathrm{L}, y$ are also largely uncorrelated with the charge $\qeff$ (see, Fig. \ref{fig:2D_lkhd_y_Q} for an illustration). Thus, to compute the expected bounds on $\qeff$, to a good approximation one needs to compute the likelihood in $\qeff$ only. Thus, we employ the following approximation of the expectation value of the likelihood
\begin{equation}
\mathcal{L}(\qeff) \equiv \langle p(d|\qeff,\mathcal{H}_\mathrm{ML}) \rangle \simeq \exp{\left[ - \rho^2 \, \mathcal{M}(\qeff_\mathrm{tr},\qeff) \right]}~. 
\label{eq:approx_likelihood}
\end{equation}
Above $\rho$ is the signal-to-noise ratio (SNR) of the signal in the data and $\mathcal{M}(\qeff_\mathrm{tr}, \qeff)$ is the \emph{mismatch} between the injected (true) waveform $h^L(f, \qeff_\mathrm{tr})$ and the waveform  $h^L(f,\qeff)$ with charge $\qeff$, such that, 
\begin{equation}
\mathcal{M}(\qeff_\mathrm{tr}, \qeff) \equiv 1 - 4 \int_{f_\mathrm{low}}^{f_\mathrm{upp}} \frac{h^L(f, \qeff_\mathrm{tr}) \, h^{L*}(f,\qeff) \, df }{S_n(f)}~. 
\end{equation}
Here ${f_\mathrm{low}}$ and ${f_\mathrm{upp}} $ denote the lower and upper frequency cutoff the detector band and $S_n(f)$ the one-sided power spectral density of the detector noise. 

\begin{figure*}[t]
\centering\
\includegraphics[width=6in]{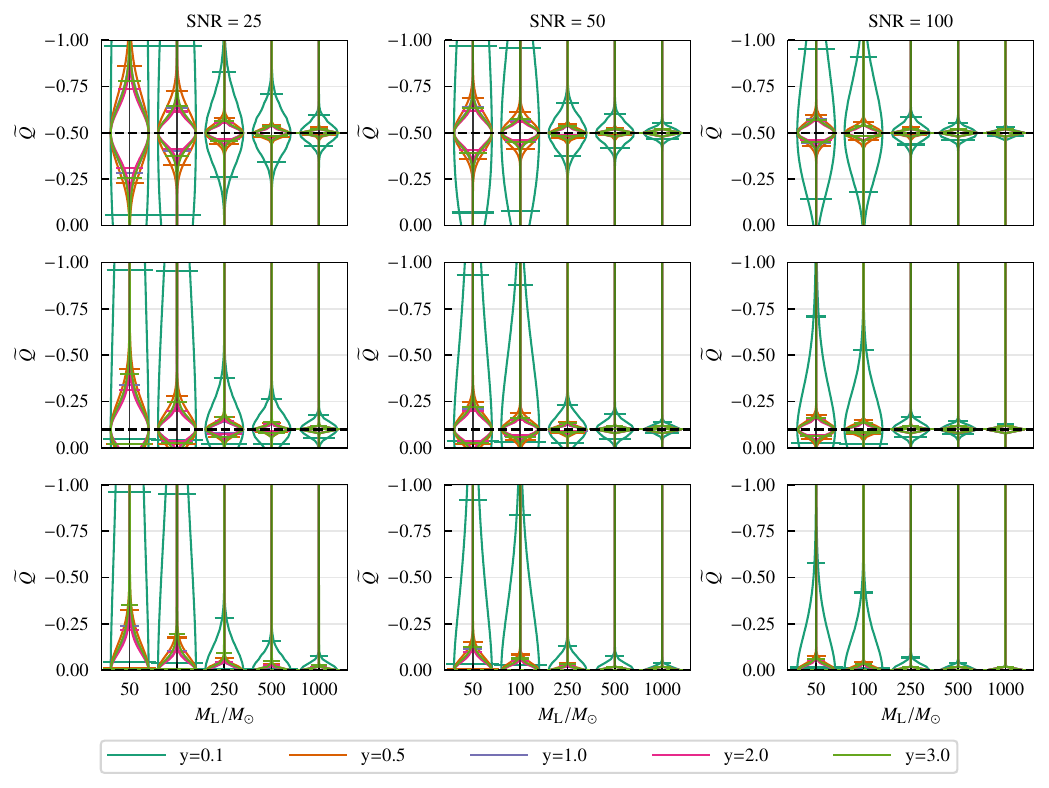}
\caption{Posteriors with $90\%$ credible bounds on $\qeff$ using aLIGO PSD, for various values the source position $y$ (shown in the legend). Panels from left to right correspond to SNR values of $25, 50 \text{ and } 100$, respectively. In the top, middle and bottom panels, the GW signal is lensed by a BH of effective charge $\qeff = -0.5, -0.1$ and $0$, respectively (indicated by horizontal dashed black lines). Here, we use the prior $\qeff \leq 0$, that corresponds to situations such as the braneworld scenario.}
\label{fig:Q90_A+negative}
\end{figure*}

\begin{figure*}[t]
\centering\
\includegraphics[width=6in]{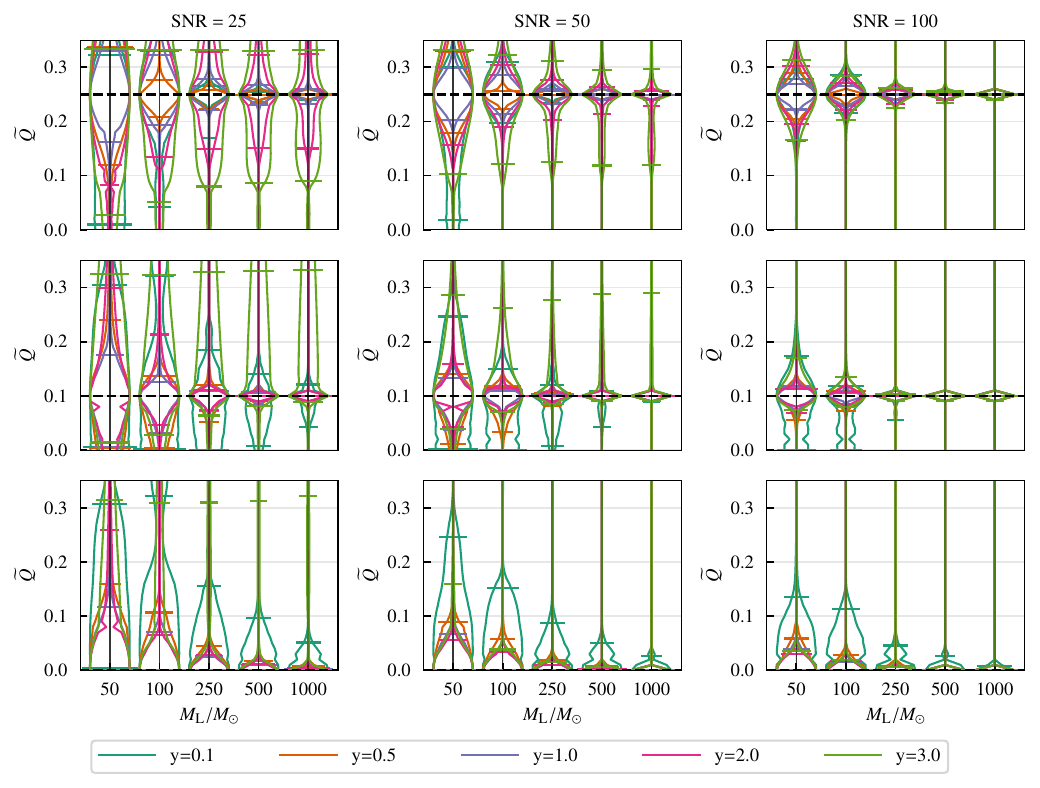}
\caption{Same as Fig.~\ref{fig:Q90_A+negative}, except that the GWs are lensed by BHs of charge $\qeff = 0.25, 0.1$ and $0$, respectively (indicated by horizontal dashed black lines). Here, we use the prior $0 \leq \qeff \leq 0.35$, which can correspond to a charge of EM or alternative gravity origin. Note that we have kept the prior such that the no-image parameter space is excluded. This is because there is no lensing of GWs in that parameter space.}
\label{fig:Q90_A+positive}
\end{figure*}

When we assume a flat prior in $\qeff$, the expectation value of the posterior distribution $P(\qeff|d)$ is the same as the likelihood $\mathcal{L}(\qeff)$. Hence we present $\mathcal{L}(\qeff)$ and its 90\% credible upper limits in Fig. \ref{fig:Q90_A+negative} and Fig. \ref{fig:Q90_A+positive}. In these figures, we have assumed the PSD of the advanced LIGO targetted for the O5 observing run (A+ configuration) \cite{LIGOScientific:2014pky}. Additionally, the unlensed signal is assumed to be due to a non-spinning equal mass BH binary with component masses $20M_\odot$ each. We compute the likelihood assuming true values of $(\qeff_\mathrm{tr}/M_{\rm L})=0, -0.1 \text{ and } -0.5$ in Fig. \ref{fig:Q90_A+negative} and for the following true values of $(\qeff_\mathrm{tr}/M_{\rm L})=0, 0.1 \text{ and } 0.25$ in Fig. \ref{fig:Q90_A+positive}. 

\begin{figure*}[t]
\centering\
\includegraphics[scale=1]{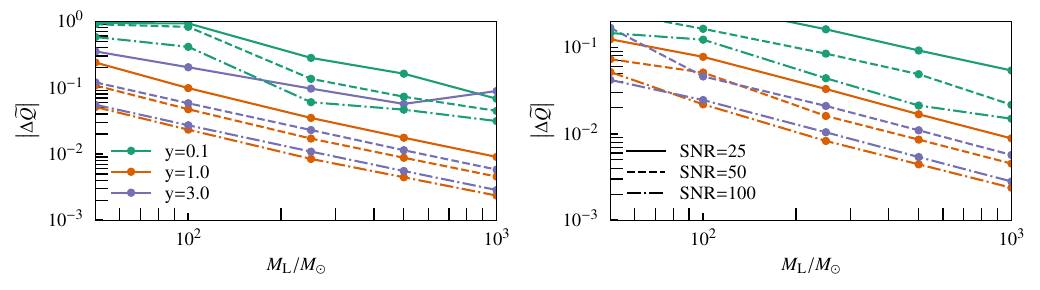}
\caption{\emph{Left panel:} The $90\%$ bounds on values of negative $\qeff$ as a function of lens mass $M_L$ for various source positions $y$ (shown by different colors) and SNR values (different line styles). \emph{Right panel:} Same as the left panel except the charge here is assumed to be $\qeff \geq 0$. This is assuming that the true lens is uncharged.}
\label{fig:summary_statistic}
\end{figure*}

It is evident from Eq.\eqref{eq:approx_likelihood} that the likelihood drops when the mismatch between the true value $\qeff_\mathrm{tr}$ and the chosen value $\qeff$ is large (i.e., when the two waveforms are very different) or when the SNR of the signal is large. These expected features can be seen in these figures. The precision in measuring $\qeff$ gets better at high SNRs and at high lens masses. Fig.~\ref{fig:summary_statistic} presents a summary of the upper limit on $|\qeff|$ when the lens is actually chargeless. Our ability to constrain various exotic scenarios will depend on these bounds. The fact that the bounds get better at high SNRs is obvious. The bounds are better at high lens masses ($M_L \sim 1000 M_\odot$) is due to the fact that these masses introduce a higher degree of wave optics lensing effects in the GW signal. As expected, the bounds are in general weaker for large values of $y$, since the lensing effects are weaker. Contrary to our na\"ive expectation, the bounds also get weaker for very small $y$, with a `sweet spot' around $y \simeq 1$.




Here we comment on some of the possible interpretations on the bound on $\qeff$. If astrophysical objects have large negative charge, then it will be a tell-tale signature of the existence of extra dimension. Thus, any bound on the negative charge can be further translated to a corresponding bound on the size of the extra dimension~\cite{Chamblin:2000ra}. This is achieved by embedding the four-dimensional brane spacetime on a higher dimensional bulk. Even though there are no analytical results available for the embedding, one can find out the extent of the extra dimension $\chi$ numerically, by integrating the Einstein's equations, together with appropriate junction conditions, along the extra dimension. For $|\mathcal{Q}/M_{\rm L}^{2}|\sim\mathcal{O}(1)$, the size of the extra dimension in the braneworld scenario is bounded from below by the ratio $(G_{5}/G_{4})$, where $G_5$ and $G_4$ are the Newton's constants in the five dimensional bulk and four dimensional brane, respectively. If $G_5$ is in the electroweak scale\footnote{Large Hadron Collider experiment has ruled out the existence of extra spatial dimension at energy scales smaller than the electroweak scale, which is generally considered to be at a few TeV. With optimal sensitivity, the constraint can increase upto ten times the electroweak scale} 
the size of the extra dimension becomes $\mathcal{O}(10^{-18}$m). Thus, for typical negative values of $\qeff$ that we will be able to constrain ($|\qeff| \lesssim 10^{-3}$), it follows that, $\mathcal{Q}/M_{\rm L}^{2}\lesssim 10^{8}$, which would bound the size of the extra dimension to $\gtrsim 0.1(G_{5}/G_{4})$\footnote{The size of the horizon extending to the fifth dimension has the following behavior: $(\mathcal{Q}/M_{\rm L}^{2})\sim \exp(\ell/\chi)$, from which we notice that if $(\chi/\ell)\sim \mathcal{O}(10^{-1})$, the dimensionless charge ratio will be $\mathcal{O}(10^{8})$. Note that the connection between the charge $\mathcal{Q}$ and the size of the extra dimension has been explored for $(\mathcal{Q}/M_{\rm L})\sim \mathcal{O}(1)$. We have assumed that a similar connection will be true for even larger values of the charge.}. This is because large negative $\mathcal{Q}$ demands smaller size for the extra dimension, and these two are exponentially related. If we take $G_5$ to be in the electroweak scale, this would provide a lower bound of $10^{-19}$m on the size of the extra dimension. This will be complimentary to the small-scale tests of Newton's law \cite{Long:2003dx, Masuda:2009vu, Sushkov:2011md, Yang:2012zzb}, which can provide upper bounds to the same.

\section{Conclusion}\label{sec:conclusion}
Upcoming observations are expected to detect gravitationally lensed GWs. One of the possible lenses is compact objects such as BHs. If the gravitational radii of these lenses are comparable to the wavelength of GWs, lensing will produce wave optics effects, producing characteristic deformations in the observed signals. The exact nature of these deformations will depend on the precise spacetime geometry (lensing potential) of the lens. Thus, lensing observations can potentially probe the detailed nature of these lensing objects. 

In this paper, we derived the lensing potential of a `charged' point mass lens. The charge $\mathcal{Q}$ can have an EM origin, in which case it is a positive definite quantity ($\mathcal{Q}$ is the square of the electric charge $q$), but can also arise in (at least) four different situations, all of which are beyond GR. These include --- (a) the braneworld scenario, where the presence of an extra spatial dimension modifies Einstein's equations, and introduces a charge term $\mathcal{Q}$ that is negative; (b) the Gauss-Bonnet theory in higher dimensions, which also leads to an effective four-dimensional spacetime with a negative charge; (c) $f(T)$ theories of gravity, as well as (d) a certain class of Horndeski theories also brings in a charge term in the spacetime metric, which is a positive quantity. Thus, negative values of the charge definitely hint at the presence of extra dimensions, while positive values of the charge can have an EM origin, or can also arise from modified theories of gravity. If one can detect the charge hair of a BH spacetime, it is possible to comment on the fundamental questions, e.g., the existence of extra dimension, as well as tell-tale signature of gravity theories beyond GR. 

Using the lensing potential of charged {point mass lens} that we derived, we computed the deformation of the lensed GW signals considering wave optics effects. This was done using a numerical scheme that we recently developed. This scheme can be used to compute the frequency-dependent lensing magnification for arbitrary lensing potentials. We noticed interesting new observables in the case of charged lenses. Owing to the axial symmetry of the lensing potential, the images and the source always lie on a line on the lens plane (same as the case of uncharged lens). For a negatively charged lens there are always two images (same as the uncharged lens), while for a positively charged lens, there can be two or four images, depending on the value of the charge and the source position. This new feature can be understood in terms of the (numerically computed) structure of the caustics of charged lenses. This introduces rich and complex effects in the lensed GW signals which are absent in those lensed by uncharged {BHs}. These additional features will help us to identify the presence of positively charged BH lenses, if they exist. On the other hand, negatively charged lenses produce features very similar to those of uncharged {BHs}, making it difficult to distinguish them from uncharged {BHs}. 

We then explored the ability of future lensing observations to constrain the charge of the lens. We consider lensing observations by a single LIGO detector with sensitivity anticipated in the O5 observing run. We showed that modest constraints on the charge can be obtained using observation of lensed GWs. Our ability to constrain the charge parameter is weakened by the fact that the charge-dependent term in the lensing potetial has a coefficient that is $\mathcal{O}(M_L/D_L)^{1/2}$, which is very small ($\sim 10^{-10}$) in typical astrophysical lensing scenarios (here, $M_L$ is the mass of the lens and $D_L$ is the distance between the lens and the observer). Even for a maximally charged BH, the effect of electric charge is too weak to be measurable in the foreseeable future. Thus, any measurable charge will be of modified gravity origin; not electric charge. If the GW lensing confirms a positive value of charge, it is very likely that the lens describes a naked singularity. On the other hand, if a negative charge is confirmed, the lens could be a BH in an alternative theory of gravity (e.g., braneworld scenario).

Note that the expected constraints that we present here employ an approximate likelihood, and neglect the possible correlations between the lensing-induced modulations in the GW signals and modulations induced by other physical effects such as orbital eccentricity of the binary source. We also ignored the possible degeneracy between a GW signal lensed by a charged BH and a GW signal lensed by an uncharged point mass lens in the presence of a macro lens (e.g., a galaxy), which can introduce more complex features. Additionally, we also neglected the effect of the spin of the BH lens. While we expect these broad conclusions to hold, the precise forecasts of the prospective constraints need to be revisited in the future considering these additional complexities. 

The numerical scheme that we employ to compute lensing magnification for arbitrary lensing potentials is too expensive to employ in actual GW parameter estimation which will require a large number of likelihood evaluations. In order to employ in GW parameter estimation, we need to develop some surrogate/semi-analytical models that interpolate the numerically computed lensing magnifications over the parameter space of interest. 

In this paper, we have worked with static and spherically symmetric spacetimes so far, and hence it would be interesting to generalize the same to rotating spacetimes, as all astrophysical objects are, in general, rotating. One could explore the possibility of measuring the spin of the compact object from lensing observations. One could also explore more general spacetimes: here we have considered the cases where $-g_{tt}=g^{rr}$; it will be useful to understand how to derive the lensing potential for spacetimes, with $-g_{tt}\neq g^{rr}$. Another possibility is probing the astrophysical environment of BHs using lensing observations. We hope to come back to these issues in future works. 

Note that some of the modifications to GR that induce an effective charge on BHs could also cause other effects in the generation and propagation of GWs, which we neglect here. Our proposal should be seen as a way of effectively checking the consistency of the GW signal that is lensed by a BH in GR (in this paper the Schwarzschild metric). Any observed inconsistency with the Schwarzschild lens will need to be investigated further in order to ascertain the nature of the charge. This is similar in spirit to various other tests of GR using GW observations. In any case, future observations of lensed GWs are very likely to offer new ways of probing the nature of compact objects.

\begin{acknowledgements}
    We are grateful to Otto Hannuksela for the careful review of the manuscript and useful comments, and to the anonymous referee for pointing out an error in the earlier version of our calculation. We thank the members of the Astrophysics \& Relativity group at ICTS for their valuable input. Our sincere thanks to Rajaram Nityananda for the productive discussions.
    We acknowledge the support of the Department of Atomic Energy, Government of India, under project no. RTI4001.
    M.A.S.’s research was, in addition, supported by the National Research Foundation of Korea
    under grant No. NRF-2021R1A2C2012473.
    S.C. thanks the Albert Einstein Institute for its warm hospitality, where a part of this work was performed. The visit to the Albert Einstein Institute is funded by the Max-Planck Society through its Max-Planck-India mobility grant. Research of S.C. is supported by the Mathematical Research Impact Centric Support (MATRICS) and the Core research grants from ANRF, SERB, Government of India (Reg. Nos. MTR/2023/000049 and CRG/2023/000934).
    Computations were performed using the Alice computing cluster at the International Centre for Theoretical Sciences. 
\end{acknowledgements}

\appendix
\section{Lensing potential for a charged lens}\label{appna}

In this appendix, we will motivate and then determine the lensing potential of a charged lens. In general, the metric of a static and spherically symmetric spacetime can be expressed as,
\begin{align}\label{eq:sphsymm}
ds^{2}=-\left[1+2U(r)\right]dt^{2}+\frac{dr^{2}}{1+2V(r)}+r^{2}d\Omega^{2}~,
\end{align}
where $U$ and $V$ are both proportional to the Newton's constant $G$ and can in general be different; $d\Omega^2\equiv d\theta^2+\sin^2\theta \,d\phi^2$, is the metric on the two-sphere. The corresponding transition to the isotropic coordinates $(R, \theta, \phi)$ reduces the above metric to the following form,
\begin{align}\label{eq:isotrop}
ds^{2}=-\left[1+2U(R)\right]dt^{2}+\left[1-2h(R)\right]\left(dR^{2}+R^{2}d\Omega^{2}\right)~,
\end{align}
where $U(R)$ and $h(R)$ are the metric functions in the isotropic coordinates. Note that the spherically symmetric radial coordinate $r$ and the isotropic radial coordinate $R$ are related as,
\begin{align}\label{rtoR}
r=R\sqrt{1-2h(R)}\sim R\left[1-h(R)\right]+\mathcal{O}(G^{2})~.
\end{align}
Since we are interested only in the weak gravity limit, we will ignore any corrections of $\mathcal{O}(G^2)$ to $U(R)$ and $h(R)$. The above relation also provides the following connection between the function $V(r)$ in Eq. (\ref{eq:sphsymm}) and the function $h(R)$ in Eq. (\ref{eq:isotrop}), such that, 
\begin{align}\label{isotropicmetric}
V(R)=-R\dfrac{dh}{dR}~.
\end{align}
For Schwarzschild spacetime, which depicts the geometry of a point mass lens, at leading order in the Newton's constant, we obtain $V(R)=-(GM_{\rm L}/R)$, and hence the above differential equation can be solved to obtain, $h(R)=V(R)=U(R)$. Note that the above equality does not hold true if terms of $\mathcal{O}(G^{2})$ are also included in the analysis.

For our purpose, we need to determine the lensing potential for the charged lens, and hence we start with the Reissner-Nordstr\"om (RN) metric in the static and spherically symmetric coordinate system:
\begin{align}
ds^{2}=&-\left(1-\frac{2GM_{\rm L}}{r}+\frac{G\mathcal{Q}}{r^2}\right)\,dt^2 
\nonumber\\
& + \left(1-\frac{2GM_{\rm L}}{r}+\frac{G\mathcal{Q}}{r^2}\right)^{-1}dr^2 + r^2 d\Omega^2 ~.
\label{eq:RN_metric}
\end{align}
Thus, to the leading order in the Newton's constant $G$ the metric function $V(r)$ reads: $V(r)=-GM_{\rm L}/r+G\mathcal{Q}/2r^{2}$. Since, $V(r)$ is already linear in the Newton's gravitational constant, we can safely set $r=R$ and obtain $V(R)$, as any correction to the $r=R$ relation is $\mathcal{O}(G)$. On substituting the metric function $V(R)$, in Eq.(\ref{isotropicmetric}), the corresponding metric function $h(R)$ in the isotropic coordinate reads, 
\begin{align}\label{hRcharge}
h(R)=-GM_L/R+G\mathcal{Q}/4R^{2}\neq V(R)~,
\end{align}
in stark contrast to the case of a point mass lens. Note that any term of $\mathcal{O}(M^{2}/R^{2})$ is necessarily of $\mathcal{O}(G^{2})$ and hence does not contribute in the weak gravity regime. Thus in the weak gravity regime, the coefficient of $(1/R^{2})$ term is solely given by the charge of the lens. This also yields the following transformation, between the radial coordinate in spherically symmetric and isotropic coordinates,
\begin{equation}
r(R) = \frac{1}{4R}\left(-G\mathcal{Q}+4GM_{\rm L}R+4R^2\right)~.
\end{equation}
It is straightforward to check that the above relation is consistent with Eqs.(\ref{rtoR}) and (\ref{hRcharge}).
Therefore, in the isotropic coordinates, the metric takes the following form (expanding in orders of $G$ and keeping only $\mathcal{O}(G)$ terms),
\begin{align}
ds^2 = & -\left(1-\frac{2GM_{\rm L}}{R}+\frac{G\mathcal{Q}}{R^2}\right) dt^2 
\nonumber\\
&+\left(1+\frac{2GM_{\rm L}}{R}-\frac{G\mathcal{Q}}{2R^2}\right)(dR^2 + R^2\,d\Omega^2)+\mathcal{O}(G^2)~.
\label{eq:RN_metric_isotropic_upto_G}
\end{align}
Comparing with Eq.\eqref{eq:isotrop}, we find the $h(R)$ to be identical to Eq.\eqref{hRcharge}, while $U(R)$ reads,
\begin{equation}
U(R) = \left(-\frac{GM_{\rm L}}{R}+\frac{G\mathcal{Q}}{2R^2}\right)~.
\end{equation}
Due to the presence of the gravitating potential, the speed of the lensed GW decreases, therefore, leading to an effective refractive index, $n$, given by,
\begin{equation}
n=\frac{1-2h(R)}{1+2U(R)}\simeq 1-\left[h+U\right]\equiv 1-2\Phi~,
\end{equation}
where we have kept only linear order terms in $G$, and have used the fact that both $h$ and $U$ are linear in $G$, and have introduced the effective Newtonian potential, 
\begin{equation}\label{eq:effective_newtonian_potential}
\Phi \equiv \frac{h+U}{2} = \frac{-GM_{\rm L}}{R} + \frac{3G\mathcal{Q}}{8R^2}~.
\end{equation}
We would like to emphasize that the result $h=U$, and hence $\Phi=U$, holds only for a point mass lens, and in general neither of these results hold true. Adopting the Born approximation, the two dimensional deflection angle associated with the above metric is found to be \cite{gravitationallenses},
\begin{equation}\label{eq:deflection_angle_definition}
\hat{\vec{\alpha}} = 2\int_{-\infty}^{\infty}\vec{\nabla}_x\Phi\, dz ~  ,
\end{equation}
where $\vec{\nabla}_x$ denotes the gradient in the lens plane. The lens is considered to be at $z=0$, with $z$ being the line-of-sight coordinate, perpendicular to the lens plane. 

Let $\vec{\xi}$ denote the Cartesian coordinates on the lens plane, with the lens at the origin. Performing the integration in Eq.\eqref{eq:deflection_angle_definition} for the gravitational potential $\Phi$, defined in Eq.\eqref{eq:effective_newtonian_potential}, we obtain,
\begin{equation}\label{eq:deflection_angle_in_xi}
\hat{\vec{\alpha}}=\left(\frac{4GM_{\rm L}}{\xi}-\frac{3\pi G Q^2}{4\xi^2}\right)\frac{\vec{\xi}}{|\vec{\xi}|}~,
\end{equation}
This result is in agreement with the result of \cite{Eiroa:2002mk} in the weak gravity limit, where terms only upto $\mathcal{O}(G)$ are retained. We now introduce the dimensionless coordinate $\vec{x}\equiv\vec{\xi}/\xi_0$, where, $\xi_0$ is an arbitrary length scale (it is set to be the Einstein radius in the main text). In addition, we introduce the scaled deflection angle \cite{gravitationallenses},
\begin{equation}\label{eq:scaled_deflection_angle_definition}
\vec{\alpha}(\vec{x})\equiv\frac{D_{\rm L}D_{\rm LS}}{\xi_0 D_{\rm S}}\hat{\vec{\alpha}}(\xi_0\vec{x})~.
\end{equation}
Using Eq. \eqref{eq:deflection_angle_in_xi}, the scaled deflection angle is found to be,
\begin{equation}\label{eq:deflection_angle_in_x}
\vec{\alpha}(\vec{x})=\frac{4 G M_{\rm L}D_{\rm L}D_{\rm LS}}{\xi_0^2D_{\rm S}}\left[\frac{1}{|\vec{x}|}-\frac{3\pi \mathcal{Q}}{16\xi_0 M_{\rm L}}\frac{1}{|\vec{x}|^2}\right]\frac{\vec{x}}{|\vec{x}|}~.
\end{equation}
Given the above two-dimensional deflection angle, we obtain the corresponding lensing potential, $\psi$ through the following relation:
\begin{equation}\label{eq:relation_deflection_angle_potential}
\vec{\nabla}_{x}\psi(\vec{x})=\vec{\alpha}(\vec{x}).
\end{equation}
Using Eq. \eqref{eq:deflection_angle_in_x} and Eq. \eqref{eq:relation_deflection_angle_potential}, we obtain the lensing potential to read,
\begin{equation}
\psi(\vec{x}) = \frac{4 G M_{\rm L}D_{\rm L}D_{\rm LS}}{\xi_0^2D_{\rm S}}\left[\ln{|\vec{x}|}+\frac{3\pi \mathcal{Q}}{16\xi_0 M_{\rm L}}\frac{1}{|\vec{x}|}\right]~.
\end{equation}
Note that the effect of the charge is suppressed by the Einstein radius $\xi_{0}$. 

\color{black}

\bibliography{References}

\end{document}